\documentstyle[preprint,aps]{revtex}
\begin{document}
\preprint{\begin{tabular}{c}
\hbox to\textwidth{April 1997 \hfill BROWN-HET-1071}\\[-10pt]
\hbox to\textwidth{ \hfill hep-ph/9704253}\\[-10pt]
\end{tabular}}
\draft
\title{ New Class of Quark Mass Matrix and Calculability of
Flavor Mixing Matrix 
\footnote{ Supported in part by the U.S. DOE Contract
DE-FG02-91ER 40688-Task A.}}
\author{Kyungsik Kang and Sin Kyu Kang\footnote{ Korea Science and Engineering
Foundation Post-doctoral Fellow.}}
\address {\it Department of Physics, Brown University, 
 Providence, RI 02912, USA} 
\maketitle
\begin{abstract}
We discuss a new general class of mass matrix ansatz that respects
the fermion mass hierarchy and {\it calculability} of the flavor mixing matrix.
This is a generalization and justification of the various specific forms 
of the mass matrix 
by successive breaking of the maximal permutation symmetry.
By confronting the experimental data, a large class of the mass 
matrices are shown to survive, while certain specific cases are 
phenomenologically ruled out. 
Also the CP-violation turns out to be maximal, when the phase
of the (1,2) element of the mass matrix is $\pi /2$.
\end{abstract}

\newpage
With the discovery of the top quark \cite{top}, 
the three family structure of the fermion 
sector has been completely determined.
Nevertheless, the flavor mixing and fermion masses and their hierarchical 
patterns remain to be one of the basic problems in particle physics.

Within the standard model, all masses and flavor mixing angle  are free 
parameters and no relations among them are provided.
Perhaps, a new theory could predict all masses and flavor mixing parameters
in terms of some new, few fundamental parameters, but we lack such
theory at the moment and are unable to derive the masses and the flavor
mixing parameters from the first principles.
One can at the best take a phenomenological standpoint in that  one assumes
 a special form for the mass matrices and hopes to be able to derive
phenomenologically viable relations for the flavor mixing parameters
in terms of the quark masses.

As an attempt to derive relationship between the quark masses and flavor mixing
hierarchies, mass matrix ansatz based on flavor democracy with a suitable 
breaking so as to allow mixing between the quarks of near kinship
was suggested about two decades ago \cite{wein}.
This in fact reflects the {\it calculability} \cite{wein,cal}
that all flavor mixing 
parameters depend solely on and are determined by the quark masses.
In general, the {\it calculability} condition does not determine the
CP violation phase, for which  either additional ansatz or input
is needed to determine.
Of several ansatz proposed, the canonical mass matrices of the 
Fritzsch type \cite{wein,test} have been generally assumed to predict the 
entire Kobayashi-Maskawa (KM) matrix \cite{km} or the Wolfenstein mixing
matrix \cite{wolf}.
Though the ansatz of the Fritzsch texture \cite{wein,cal} is attractive 
because of its maximal {\it calculability},
it predicts a top quark mass to be no larger than 100 GeV and thus is 
ruled out \cite{test}. 

Alternatively one may introduce a modification to the Fritzsch texture of mass
matrix by allowing a non-vanishing (2,2) elements in the ``hierarchical"
mass eigenstates.
Such scheme was proposed sometimes ago by Kaus and Meshkov
\cite{kaus} based on a postulate
of the ``BCS mechanism" for the quarks and assuming that the heaviest
third generation quark mass is to be identified by the non-zero eigenvalue
of the ``democratic mass matrix".
More recently Fritzsch {\it et al}. \cite{fr2} have suggested the same type of
mass matrix by assuming that the ``democratic" maximal permutation symmetry
may be broken in a simple and analogous manner as
the mass mixing pattern of the $\eta $-$\eta^{\prime}$ system. 
As a result the mass matrices contain only three zero elements at
(1,1), (1,3) and (3,1) position in the hierarchical mass eigenstates.
Nevertheless, this does not necessarily
imply  lack of  {\it calculability} because the additional 
non-vanishing (2,2) element may be related 
to the (2,3) or (3,2) elements.

With one such form for the mass matrices, 
Fritzsch {\it et al}.\cite{fr2}  described the KM 
matrix in terms of the quark mass ratios to the lowest order approximation 
and claimed that they are in good agreement with the experimental values.
However, this is not true at least for $V_{cb}$ element 
because one gets $|V_{cb}| \simeq \frac{1}{\sqrt{2}} \left( \frac{m_s}{m_b}
-\frac{m_c}{m_t}\right) $ so that $m_t(\mu=1~\mbox{GeV})$ 
can be at most 113 GeV 
from the experimental range $|V_{cb}|=0.036 - 0.046 $ \cite{data}.
Several other authors \cite{wu,xing} have also discussed specific
forms of this type of mass matrices.

We present in this paper a generalization of this class of mass matrices 
in such a way 
that it can maintain the {\it calculability} property and consistency with 
experiments, while accommodating a CP violation phase.
We will show that this can be achieved by breaking the democratic flavor
symmetry $S(3)_L \times S(3)_R $ successively down to
$S(2)_L \times S(2)_R$ and to $S(1)_L \times S(1)_R$, so that the (2,2)
element can be related to (2,3) element appropriately in the hierarchical
mass eigenstates.

As is well known, the $3\times 3$ ``democratic mass matrix ",
\begin{eqnarray}
   \frac{c}{3}~\left( \begin{array}{ccc}
               1 & 1 & 1 \\
               1 & 1 & 1 \\
               1 & 1 & 1  \end{array} \right),
\end{eqnarray}
exhibits the maximal $ S(3)_{L} \times S(3)_{R} $ permutation symmetry.
This can be achieved by breaking of the chiral symmetry $U(3)_L \times
U(3)_R$ to $S(3)_L \times S(3)_R$, where $U(3)$ is the symmetry group 
connecting the three generations \cite{kaus,fr3}.
One may say that the scale of this chiral symmetry breaking is the electroweak
symmetry breaking scale at which the third generation quarks get heavy masses.
Indeed, one can see this by making
unitary transformation of (1) with the help of 
\begin{equation}
   U = \left( \begin{array}{ccc}
               \frac{1}{\sqrt{2}} & -\frac{1}{\sqrt{2}} & 0 \\
               \frac{1}{\sqrt{6}} & \frac{1}{\sqrt{6}} & -\frac{2}{\sqrt{6}} \\
               \frac{1}{\sqrt{3}} & \frac{1}{\sqrt{3}} & \frac{1}{\sqrt{3}} 
               \end{array} \right).
\end{equation}
This matrix is in fact reminiscent of the matrix for the mass squared of
the neutral pseudoscalar mesons in QCD in the chiral limit.
In order to account for the hierarchical pattern of the second and first
generation quark masses,  one has to break the $S(3)_{L} \times S(3)_{R} $ 
symmetry successively in two stages to $S(2)_L \times S(2)_R$ and
$S(1)_L \times S(1)_R$.
This can be achieved by adding the following two matrices to the ``democratic 
matrix" (1):
\begin{eqnarray}
               \left( \begin{array}{ccc}
               0 & 0 & a \\
               0 & 0 & a \\
               a & a & b \end{array} \right),
 ~~ \qquad
         d ~\left( \begin{array}{ccc}
               1 & 0 & -1 \\
               0 & -1 & 1 \\
              -1 & 1 & 0  \end{array} \right),
\end{eqnarray}
where the parameters $(a,b)$ and $d $ are responsible for the breakdown
of $S(3)_{L}\times S(3)_{R}$ and $S(2)_{L}\times S(2)_{R}$ symmetries,
respectively.
It is also reasonable to anticipate that this two-stage breaking happens to 
be at around 1 GeV
scale in view of the proximity of the second and first generation quark
masses compared to the third generation quarks.
Note that the two scales in proximity are related to the generation of 
the second and first generation quark masses and
the evolution from the electroweak scale to 1 GeV scale can not
alter the `` democratic " pattern of the mass matrix because of the 
symmetry $S(3)_L \times S(3)_R$.
Thus the resulting mass matrix can be regarded as the one at 1 GeV scale.

In principle, the most general form of 
$S(3)_{L}\times S(3)_{R} \rightarrow S(2)_{L}\times S(2)_R $ breaking can
allow different parameters at the (1,3), (2,3), (3,1) and (3,2) elements.
But to maintain the {\it calculability} property,
the form of (3) containing only two parameters $(a,b)$ is necessary,
which is general enough to cover all different specific forms
proposed by others \cite{kaus,fr2,wu,xing} as a special case.
Then in the hierarchical basis after the unitary transformation with (2), 
the resulting mass matrix $M_H$ becomes
\begin{eqnarray}
   M_H = \left( \begin{array}{ccc}
               0 & A & 0 \\
               A & D & B \\
               0 & B & C  \end{array} \right),
& ~~ \qquad
\end{eqnarray}
where $A=\sqrt{3}d, D=-\frac{2}{3}(2a-b), B=-\frac{\sqrt{2}}{3}(a+b)$ and
$C= \frac{1}{3}(4a+b)+c $.  

Note that in order to get a hermitian mass matrix instead of (4), one can use
the following two matrices
\begin{eqnarray}
          \left( \begin{array}{ccc}
            p & p  & a+q \\
            p & p  & a+q \\
            a+q^{\ast} & a+q^{\ast} & b-2p 
               \end{array} \right), 
 ~~ \qquad
         d~\left( \begin{array}{ccc}
               \cos \sigma & -i\sin \sigma & -e^{-i\sigma} \\
               i\sin \sigma & -\cos \sigma &  e^{-i\sigma} \\
               -e^{i\sigma} & e^{i\sigma} & 0 
                \end{array} \right),
\end{eqnarray}
where $p=\frac{4}{9}(a+b)\sin^2 \frac{\delta}{2} $ and $q=p(1+i\frac{3}{2}
\frac{e^{i\delta/2}}{\sin \frac{\delta}{2}}) $,
in such a way that the matrices (3) are  replaced by those of (5). 
Then, after the unitary transformation with (2),
the (1,2) and (2,3) elements become $A e^{-i\sigma}$ and
$B e^{-i\delta}$ respectively.
However, since  only one phase factor is sufficient to describe the 
CP-violation in the standard model containing three family generations of 
quarks, we may 
introduce only one phase factor in the  
hermitian matrix $M_H$ such that only (1,2) and (2,1) elements
are complex and conjugate to each other. 
In this way, a hermitian mass matrix of the type (4), with a complex
element at (1,2) and its conjugate at (2,1), 
can be obtained from a general permutation symmetry breaking chain, i.e.,
$S(3)_{L}\times S(3)_{R} \rightarrow  S(2)_{L}\times S(2)_{R} 
\rightarrow S(1)_{L} \times S(1)_{R} $.

At a glance, the matrix $M_H$ contains four independent parameters even in
the case of real parameters so that the {\it calculability} is lost.
However, one can make additional ansatz to relate $a$ to $b$, 
so that $a=kb$ in general, with the same ratio parameter $k$ for both the 
up- and down-quark sectors, so as to maintain the {\it calculability}.
On the other hand, one can expect that any choices other than $a=kb$ for both
up and down quarks might be interesting, but
the choice of $a=kb$ meets clearly the elegance of simplicity.
Then, the (2,2) element is related to (2,3) element by
$w\equiv B/D = (k+1)/\sqrt{2}(2k-1) $ in the
hierarchical mass eigenstate and various specific mass matrices proposed 
by others can be identified as a special case of different ratios
i.e., $w=\frac{5}{3}~(k=0.9)$ for Ref. \cite{kaus},
$w=-\frac{1}{\sqrt{2}}~(k=0)$ for Fritzsch {\it et al}. \cite{fr2}, 
$w=\pm 2\sqrt{2}~(k=\frac{5}{7} ~\mbox{or}~ \frac{1}{3}) $ for 
Ref. \cite{wu} and $w=\sqrt{2}~(k=1)$ for Ref. \cite{xing}.
The case of $k=\frac{1}{2} $ reduces to the old Fritzsch type with
$D=0 $ which is ruled out by the experiments as we said before.
We are therefore interested in the general case but $k\neq \frac{1}{2} $ 
in this paper.

The next step is then  to constrain $k$ for the general class of  mass matrix 
by confronting the experiments for consistency. 
Obviously a careful analysis with exact flavor mixing elements predicted
from the new ansatz is desired to confront the experiments.
The mass matrix $M_H$ of the type (4) can be brought to a diagonal form by 
appropriate
rotation of the fermion fields in the hierarchical eigenstates 
via a biunitary transformation,
\begin{eqnarray*}
U^{(u)}_L M^{(u)}_{H} U^{(u)^{\dagger}}_R = diag[m_u, m_c, m_t] \\
U^{(d)}_L M^{(d)}_{H} U^{(d)^{\dagger}}_R = diag[m_d, m_s, m_b] ,
\end{eqnarray*}
and the  quark fields in the physical mass eigenstates
are related to the hierarchical mass eigenstates by
\begin{eqnarray*}
q^{(u)}_{L(R)}=U^{(u)}_{L(R)}u^0_{L(R)} \\
q^{(d)}_{L(R)}=U^{(d)}_{L(R)}d^0_{L(R)}
\end{eqnarray*}
where $(q^{u}, q^{d})$ and $(u^0, d^0)$ denote the physical mass eigenstates 
and the hierarchical mass eigenstates for the up- and
down-quark sectors respectively.
We note that a phase factor is attached to the (1,2) and (2,1)
elements as $Ae^{-i\sigma}$ and $Ae^{i\sigma }$ in $M_H$, where
$A$ will be assumed to be positive without loss of generality.
Then both $U_L M_H U_L^{\dagger}$ and
$U_R M_H U_R^{\dagger}$ are diagonal so that $U_LU_R^{\dagger}\equiv K$ is 
again diagonal.
In our ansatz, it turns out in general that, because of the empirical mass
hierarchy $m_1 \ll m_2 \ll m_3 $,
$K=diag[1,-1,1]$ irrespective
of the sign of $D$ and $K=diag[-1,1,1]$ only for positive $D$. 
This point was not clearly understood in previous works 
\cite{fr2,wu,xing,gupta}.
Fritzsch {\it et al.} \cite{fr2} chose the relative signs
of the $S(3)_L \times S(3)_R $ breaking terms different,
so that the sign of $m_c$ is opposite to that of $m_s$,
while keeping $m_u$ and $m_d$ to be negative as mass
eigenvalues, which is clearly inconsistent with the
empirical quark mass hierarchy.
Other authors \cite{gupta} assumed the same form of the mass matrix
without basing on a symmetry consideration and thus
treating the up- and down-quark sectors unevenly.

The parameters $A,B,C$ and $D$ can be expressed in terms of the quark
masses.
As emphasize earlier, in this paper, we will deal with the same pattern
for both the up- and down-quark mass matrices so that the calculability of the 
flavor mixing matrix from the quark masses retained.
In view of the hierarchical pattern of the quark masses, it is natural
to expect that $A < |D| \ll C $, and then the case of $K=diag[1,-1,1]$
for positive $D$ can be excluded if the same ratio parameter $w$ is required
for both up- and down-quark sectors.
Otherwise, the masses of the second family could be unacceptably large.

{\it The Case $K=diag[-1, 1, 1]$}:  
Because a hermitian matrix can be expressed as a unitary transformation
of a real symmetric matrix, one can write 
$M^{(u,d)}_{H} = P^{(u,d)} M_r^{(u,d)} {\tilde P^{(u.d)}}$, where
$P^{(u,d)}=diag[\exp(-i\sigma^{(u,d)}), 1, 1]$,
and the real matrix $M_r^{(u,d)}$ can be diagonalized by a real
orthogonal matrix $R^{(u,d)}$ so that
$R^{(u,d)}M_r^{(u,d)}{\tilde R^{(u,d)}} = 
diag[-m_{(u,d)}, m_{(c,s)}, m_{(t,b)}].$
Then $U^{(u)}_L = {\tilde R^{(u)}}P^{(u)^{\dagger}}$ 
and $U^{(d)}_L = {\tilde R^{(d)}}P^{(d)^{\dagger}}.$
The flavor mixing matrix is given by $V=U^{(u)}_L U^{(d)^{\dagger}}_{L}=
{\tilde R^{(u)}}P^{(u)^{\dagger}}P^{(d)}R^{(d)} = {\tilde R^{(u)}} P R^{(d)}$
where $P=diag [e^{i\sigma }, 1, 1]$ with 
$\sigma = \sigma ^{(u)} - \sigma^{(d)}$.

From the characteristic equation for the $M_r$, the mass matrix $M_r$
can be written by
\begin{eqnarray}
   M_r = \left( \begin{array}{ccc}
               0 & \sqrt{\frac{m_1 m_2}{1-\frac{\epsilon}{m_3}}} & 0 \\
               \sqrt{\frac{m_1 m_2}{1-\frac{\epsilon}{m_3}}} & 
               m_2-m_1+\epsilon & w(m_2-m_1+\epsilon) \\
               0 & w(m_2-m_1+\epsilon ) &
               m_3-\epsilon  \end{array} \right)
\end{eqnarray}
in which the small parameter $\epsilon$ is related to $w$, i.e.,
$w \simeq \pm \frac{\sqrt{\epsilon m_3}}{m_2}
\left(1+\frac{m_1}{m_2}-\frac{m_2}{2m_3}\right)$,
whose range is to be determined from the experiments.
Note the sign of $B$ is undetermined from the characteristic
equation but the KM matrix elements are independent of the sign
of $B$.
Then, we can obtain analytic expressions for the flavor mixing matrix
$V$ in the leading approximation such as
\begin{eqnarray}
|V_{us}| &\simeq &\left| \sqrt{m_d/m_s}\exp{(i\sigma)}-
                        \sqrt{m_u/m_c}\right|, \\
|V_{cb}| &\simeq & \left| w
           \left(m_s/m_b-m_c/m_t\right)\right|, \\
|V_{ub}|/|V_{cb}| &\simeq & \sqrt{m_u/m_c}, ~~~
|V_{td}|/|V_{ts}| \simeq \sqrt{m_d/m_s}.
\end{eqnarray}
Notice  that $|V_{cb}|$ depends on the quark mass ratios and $w$.
In fact the $w$-dependence appears in the four elements
$V_{ub}, V_{cb}, V_{ts}$ and $V_{td}$ only.
Since the second term of $|V_{cb}|$ is negligible compared to the first term,
it is easy to examine the range of $w$ for which
$|V_{cb}|$ is compatible with experiments.
We use the light quark masses \cite{mass1},
$ m_u=5.1 \pm 0.9 ~\mbox{MeV},~ m_d=9.3 \pm 1.4 ~\mbox{MeV}~\mbox{and}~
 m_s=175 \pm 25 ~\mbox{MeV}, $
and the heavy quark masses \cite{mass2},
$ m_c=1.35 \pm 0.05 ~\mbox{GeV}~\mbox{and}~
 m_b=5.3 \pm 0.1 ~\mbox{GeV},$
all of which correspond to the masses at a $\overline{MS}$ renormalization 
point of 1 GeV.
The top quark mass $m_t$ of the recent measurement
$m_t=175 \pm 6 $ GeV corresponds to the running mass 
$m_t(\mu = 1~\mbox{GeV}) \simeq 280 - 450 $ GeV
for $\Lambda_{\overline{MS}}=150 - 200$ MeV \cite{mass3}.

Using the value $|V_{cb}| = 0.036 - 0.046 $ from experiments \cite{data},
Eq. (8) leads to $1.01\lesssim |w| \lesssim 2.02$ so that
$0.82\lesssim k \lesssim 1.31$ if $w>0$ and
$0.11\lesssim k \lesssim 0.28$ if $w<0$
in the leading approximation, which is close to the exact result
 $0.97\lesssim |w| \lesssim 1.87$ so that
$0.85\lesssim k \lesssim 1.36$ if $w>0$ and
$0.10\lesssim k \lesssim 0.26$ if $w<0$.
Note that $\epsilon \simeq O(m_1)$ for the allowed range of $k$ and $w$.

Next, we examine if this range of $w$ preserves the consistency
with experiments for other KM elements.
Since several KM elements depend on the phase factor $\sigma$,
we have to determine the allowed range of the phase factor first.  
We see from  Eq. (7) that $|V_{us}|$ depends 
on the phase factor $\sigma $, while independent of $w$.
Using the experimental value $|V_{us}| \simeq 0.219 - 0.224$ \cite{data} 
the allowed range of $\sigma $ turns out to be 
$26^{\circ}-111^{\circ}$. 
In particular,  the maximal weak CP phase conjecture
$\sigma = \pi /2 $ suggested previously by Ref. \cite{shin}
follows when $m_s \simeq 0.206$ GeV from Eq. (7) and
when $m_s \simeq 0.194$ from exact calculation 
for the central values of the parameters $m_u, m_d, m_c$ and
$|V_{us}|$.
The exact numerical result gives
$39^{\circ} \lesssim \sigma \lesssim 117^{\circ}$.
In addition we find that all other KM elements are in good agreement with
experiments for the above ranges of $w$ and $\sigma $.

{\it The Case $K=diag[1, -1, 1]$}:  For a negative $D$, the real symmetric 
matrix $M_r^{(u,d)}$ can be diagonalized as 
$R^{(u,d)}M_r^{(u,d)}{\tilde R^{(u,d)}} 
= diag[m_{(u,d)}, -m_{(c,s)}, m_{(t,b)}]$, 
thus reversing the signs of $m_1$ and $m_2$ in Eq. (6). 
As we noted, a positive $D$ in this case is excluded for the reasons of 
naturalness due to the quark mass hierarchy and {\it calculability}.
Following the similar analysis as in the previous case, we get 
$1.14\lesssim |w| \lesssim 2.76$ so that
$0.72\lesssim k \lesssim 1.17$ if $w>0$ and
$0.14\lesssim k \lesssim 0.33$ if $w<0$,
and the same range of $\sigma$ as in the previous case
in the exact numerical calculation,
while we find the same result of $w$ and $\sigma$
as in the previous case in the leading approximation.
Consequently
the ansatz adopted by Fritzsch {\it et al}. \cite{fr2},
corresponding to $k=0$,
is not consistent with
experimental data of $V_{cb}$ and the ansatz adopted by
Ref. \cite{wu}, corresponding to $w^2=8$, is 
slightly beyond the upper bound of the allowed $w$.

Now, we note that the predicted ratio $|V_{ub}|/|V_{cb}|$ ($\lesssim 0.07$)
tends to be on the low side of (but consistent with ) the present experimental 
range, 
$|V_{ub}|/|V_{cb}|=0.08\pm 0.02$ \cite{data} or $0.08 \pm 0.016$ \cite{ratio1}.

In terms of the three inner angles of the unitarity triangle \cite{uni}
$    \alpha = arg \left(-\frac{V_{ub}^{\ast}V_{ud}}
                      {V_{tb}^{\ast}V_{td}}\right),~~
    \beta  = arg \left(-\frac{V_{tb}^{\ast}V_{td}}
                      {V_{cb}^{\ast}V_{cd}}\right)~\mbox{and}~~
    \gamma  = arg \left(-\frac{V_{cb}^{\ast}V_{cd}}
                      {V_{ub}^{\ast}V_{ud}}\right),$
we obtain
$     \alpha  \simeq  \sigma \simeq  26^{0} - 111^{0}, ~
     \beta   \simeq  
        6^{0} - 13^{0}~\mbox{and}~
     \gamma  \simeq  180^{0}-\alpha - \beta  \simeq  148^{0} - 56^{0}$
in the leading order approximation.
These angles are independent of $k$.
From the Jarlskog determinant \cite{jar},
$\mbox{Det}~C\simeq \frac{(k+1)^2}{(2k-1)^2}\sqrt{m_u m_d m_s m_c}
 m_t^2 m_s^2 \sin \sigma $,
we see that the CP violation becomes maximal for $\sigma =90^{\circ}$
which is an allowed value from our results.

Finally, the Wolfenstein parameters \cite{wolf} can be determined from
$|V_{us}| \simeq \lambda, |V_{cb}| \simeq \lambda^2A,
V_{ub} \simeq \lambda^3A(\rho - i\eta )$ and
$V_{td} \simeq \lambda^3A(1-\rho-i\eta)$ in terms of the quark masses.
Since $\lambda \approx \sqrt{\frac{m_d}{m_s}}
\left(1-\sqrt{\frac{m_sm_u}{m_dm_c}}\cos \sigma \right) \simeq
|V_{12}|$  from Eq. (7), we obtain $\sigma \simeq 80.17^{\circ}$ for
all central values of quark masses and $\lambda = 0.22$.
From the element $V_{cb}$, we get $A\simeq 0.74-0.95$ and
since $|V_{ub}|/|V_{cb}|=\lambda \sqrt{\rho^2 + \eta ^2}$ ranges
$0.06 - 0.10$ from the semileptonic B decays \cite{data},
we get $\rho^2 + \eta^2 \simeq 0.074-0.207$ for the central values of 
quark masses, while from Eq.(9), $\rho^2 + \eta^2 =0.0781$.

In conclusion, we suggested a general class of hermitian mass matrices
that can be obtained from successive breaking chain
$U(3)_L\times U(3)_R \rightarrow
S(3)_L\times S(3)_R \rightarrow
S(2)_L\times S(2)_R \rightarrow
S(1)_L\times S(1)_R$
so as to reflect the quark mass hierarchy and to maintain the 
{\it calculability} of the flavor mixing matrix and its consistency with
experiments.
There are four regions of $k$, the ratio parameter of the two elements
of the $S(2)_L \times S(2)_R$ symmetric matrix, for which the
generalized mass matrix ansatz is compatible with experiments.
In particular, the CP violation turns out to be maximal when
the phase of the mass matrix is $\pi/2$.
\acknowledgements
One of us (SKK) would like to thank the Korea Science and Engineering
Foundation for a Post-doctoral Fellowship and also the members of the
High Energy Theory Group for the warm hospitality extended to him
at Brown University.

\end{document}